\documentclass[11pt,twoside]{article}

\usepackage{asp2006_hotcool}
\usepackage{epsf}
\usepackage{lscape}

\usepackage{graphicx}

\begin{document}
\title{Model Atmospheres for Cool Massive Stars}  
\author{Bertrand Plez}   
\affil{Universit\'e Montpellier 2, CNRS, GRAAL, F-34095 Montpellier, France}    

\begin{abstract}
In this review given at the Hot and Cool: Bridging Gaps in Massive Star Evolution
conference, I present the state of the art in red supergiant star atmosphere modelling.
The last generation of hydrostatic 1D LTE MARCS models publicly released in 2008 
have allowed great achievements in the past years, like the calibration
of effective temperature scales. I rapidly describe this release, and then 
I discuss in some length the impact of the opacity
sampling approximation on the thermal structure of models and on their emergent spectra.
I also insist on limitations inherent to these models. Estimates of collisional and
radiative time scales for electronic transitions in e.g. TiO suggest that non-LTE
effects are important, and should be further investigated. Classical 1D models
are not capable either to provide the large and non-gaussian velocity fields we 
know exist in red supergiants atmospheres. I therefore also present current efforts 
in 3D radiative hydrodynamical simulation of RSGs. I show that line profiles and shifts 
are predicted by these simulations, without the need for fudge
micro- and macroturbulence velocities. This is a great progress, although
 line depths and widths are slightly too shallow.
This is probably caused by the simplified grey radiative transfer used in these
heavy simulations. Future non-grey 3D simulations should provide a better fit 
to observations in terms of line strengths and widths. 
\end{abstract}

\section{Classical Model Atmospheres}
Model stellar atmospheres constitute the basis on which we interpret stellar
spectra. Our ability to produce "good" models including the necessary physical
approximations and input data directly impacts the quality and reliability
of the parameters we extract from observations: $T_{\rm eff}$, chemical 
abundances, etc.
We know red supergiant (RSG) atmospheres are subject to strong convective
motions, resulting in temperature inhomogeneities, and velocity 
fields. Particle densities are low and non-LTE effects are expected (see below).
Despite this complexity, simple one-dimensional, hydrostatic, LTE models
have been constructed, and used to study RSGs with success.
I start by presenting the current MARCS generation of such models, 
detailing some aspects linked to the opacity sampling approximation.
I then show how velocity fields affect the spectra, 
and discuss possible non-LTE effects on molecules.
Finally I present current efforts in hydrodynamical modelling, with 
encouraging results, but with their own limitations, hopefully alleviated
in the near future.
\subsection{MARCS 2008}
The MARCS model atmosphere code has been in use since the mid-70's.
The official birth certificate of MARCS, a code for Model Atmospheres in Radiative and 
Convective Scheme, is  \citet{1975A&A....42..407G}. 
It allowed the computation of hydrostatic, plane-parallel (PP),
line-blanketed atmospheres, with convection included following
\citet*{1965ApJ...142..841H} recipe for MLT, and line opacity treated in the form of Opacity Distribution Functions (ODF). Many updates were implemented 
since then, and a major release was published recently \citep{2008A&A...486..951G}.
MARCS 2008 is characterised by, e.g., new opacities for H$_2$O, atomic collisional 
line broadening
included using the description of \citet*{1995MNRAS.276..859A},
and hydrogen lines modelled
using a code by Barklem, described in \citet{2003IAUS..210P.E28B}. About
108\,000 opacity sampling points are used (see next section). All atomic and molecular line opacities were reviewed, as 
well as continuous opacities. More than half a billion lines are included, e.g. TiO,
ZrO, VO, CO, CN, MgH, just to list a few.
Full details are provided in \citet{2008A&A...486..951G}, that also relates 
some of the historical 
background, and discusses in depth the physical assumptions, numerical methods,
and physical data used. Additional historical details are provided in \citet{2008PhST..133a4003P}. About 30\,000 models have been computed at the time
 I am writing these lines in early 2009. A standard grid is available on the web
  (marcs.astro.uu.se).

\subsection{Sampling of Opacities, Model Thermal Structure, and Fluxes}
In MARCS, as well as in most modern model atmosphere codes, the opacity
is treated with the opacity sampling (OS) approximation. This is a Monte-Carlo 
evaluation of the radiation field using a set of wavelengths where the opacities,
 and monochromatic contributions to the intensities, flux, radiation 
pressure, and all radiation field characteristics are  calculated. A simple summation over wavelength gives the wavelength 
integrated quantities.  
In principle, if the number of wavelengths is large enough, this approach 
is safe. There are however two questions to be investigated: 
(i) how many OS points are needed for a given convergence 
of the model (e.g. temperature corrections $\Delta T < 1$K)?
(ii) how well is the spectrum represented, e.g. to compute synthetic photometry or 
do spectral classification?
This was discussed by \citet{2008PhST..133a4003P}, and I will only recall here what 
concerns more specifically RSGs. In these cool star atmospheres
the dominant opacity is that of molecules: CO, TiO, H$_2$O. The latter two
have a very dense spectrum with lines mostly blended with one another. This 
makes the OS approximation very well functioning. On the contrary CO, 
in the H, K and L bands has fewer, well separated lines. At a resolution of 
20\,000, the wavelength sampling does not represent evenly strong and weak 
lines, and the continuum. This is demonstrated by statistics on 
ensembles of 
models computed with various samplings: the nominal 
$R=\lambda/\Delta\lambda=20000$, 
and models with reduced $R=6700$, $R=2000$, and $R=670$, for a wavelength range between 900\AA\ and 20$\mu$m. I computed the standard deviation 
to the reference high resolution model
of models with sparser sampling.
For cool models representative of RSGs, the deviations in the temperature
structure are very small: always less than about 10K at the lowest resolution,
and less than 3K for the $R=6700$ models. This is because the opacity that 
matters for the thermal structure of these star atmospheres (TiO and H$_2$O) 
is statistically well taken into account even at low resolution. This is not 
the case for the rendering of the spectrum. Under-sampling causes local errors 
in fluxes. For RSGs, and after smoothing to $R=200$,
 errors amount to about 10\% in the blue-UV, and
about 5\% in the IR CO bands, for a sampling at $R=6700$. Sampled spectral energy
distributions (SEDs) should not be compared to observed spectra at medium-high
resolution. Even when the SEDs are degraded to low resolution (a few 100), 
systematic errors remain
at a level of several \% for an initial sampling of $R=20000$. 

In conclusion, with a sampling of wavelengths at $R=20000$, the thermal structure
of red supergiant atmospheres is accurately computed, but there may be residual 
systematic errors in the sampled SED. So, that the spectrum is somewhat wrong does
not mean that the thermal structure is! A better spectrum may be calculated
using the computed atmospheric structure and a synthetic spectrum code.
We will provide the detailed spectra either on the marcs.astro.uu.se or on the
Pollux synthetic spectra database (\cite{2008asvo.proc..217P}; http://pollux.graal.univ-montp2.fr). Finally, remember that all this is within 
the adopted approximations of LTE, hydrostatic equilibrium, and spherical 
symmetry. Additional systematic errors are expected due to real stars not behaving
in this simple way!

\section{Affects of Velocity Fields on Spectra}
Red supergiant atmospheres are strongly affected by convective motions, 
and large velocities have been measured through line shifts and broadening
\citep[e.g.][]{2007A&A...469..671J,2008AJ....135.1450G}. Two types of 
non-thermal Doppler broadening must be taken into account: microturbulence 
and macroturbulence broadening. Both reflect our ignorance of real velocity
 fields in stellar atmospheres. Microturbulence was historically introduced to 
allow for a unique abundance when using strong and weak lines of the same 
species in spectroscopic analyses. Advocating some turbulent velocity field 
at a scale smaller than the photon mean free path, one can desaturate strong lines
and increase their equivalent width at a given abundance. This was shown later 
to stem from convective motions in the case of the Sun, using sophisticated
3D hydrodynamical simulations \citep{2000A&A...359..729A}. 
Macroturbulence, a velocity
field on a larger scale, may be necessary 
in addition to allow a fit of line widths, without impacting their
equivalent width. Both microturbulent and macroturbulent velocities
are most often assumed to follow a gaussian distribution. In the case of the Sun
it was shown not to be correct: real lines are asymmetric. It is also the case
for other stars, but we do not have as sophisticated models yet, nor the 
possibility to secure as detailed observations.

It is nevertheless very interesting to scrutinise a RSG spectrum to 
try to estimate the micro- and macroturbulent parameters.
Fig. 1 and 2 show the detail of the TiO $\gamma$' 0-0 band-head
\begin{figure}[!ht]
\plotone{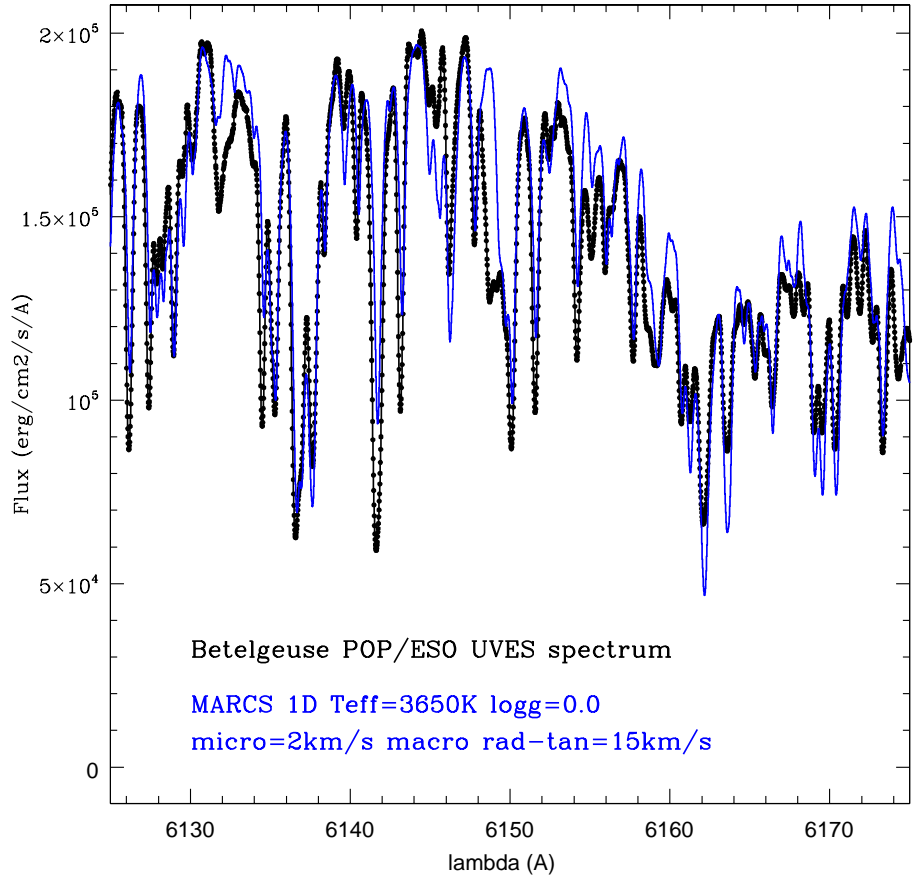}
\caption{Comparison of a calculated spectrum (thin blue line)
with a high resolution observed spectrum of Betelgeuse (black line with dots)
in the vicinity of the $\gamma$' 0-0 band-head. Most 
spectral features are satisfactorily reproduced, although not perfectly. 
Considering the number of TiO lines present in this interval, most of which
are not observed in the laboratory, and were predicted \citep{1998A&A...337..495P},
this is a very good fit. A 
microturbulence velocity of 2km/s and a
 radial-tangential macroturbulence of 15km/s were adopted.}
\end{figure}
in the spectrum of Betelgeuse \citep[POP/ESO archive,][]{2003Msngr.114...10B}.
Model spectra were computed for appropriate parameters for $\alpha$ Ori
\citep{2005ApJ...628..973L},
and then convolved with different macroturbulent velocity distributions, all
with a width of 15km/s: Gaussian, exponential, and radial-tangential
\citep{2008oasp.book.....G}.
\begin{figure}[!ht]
\plotone{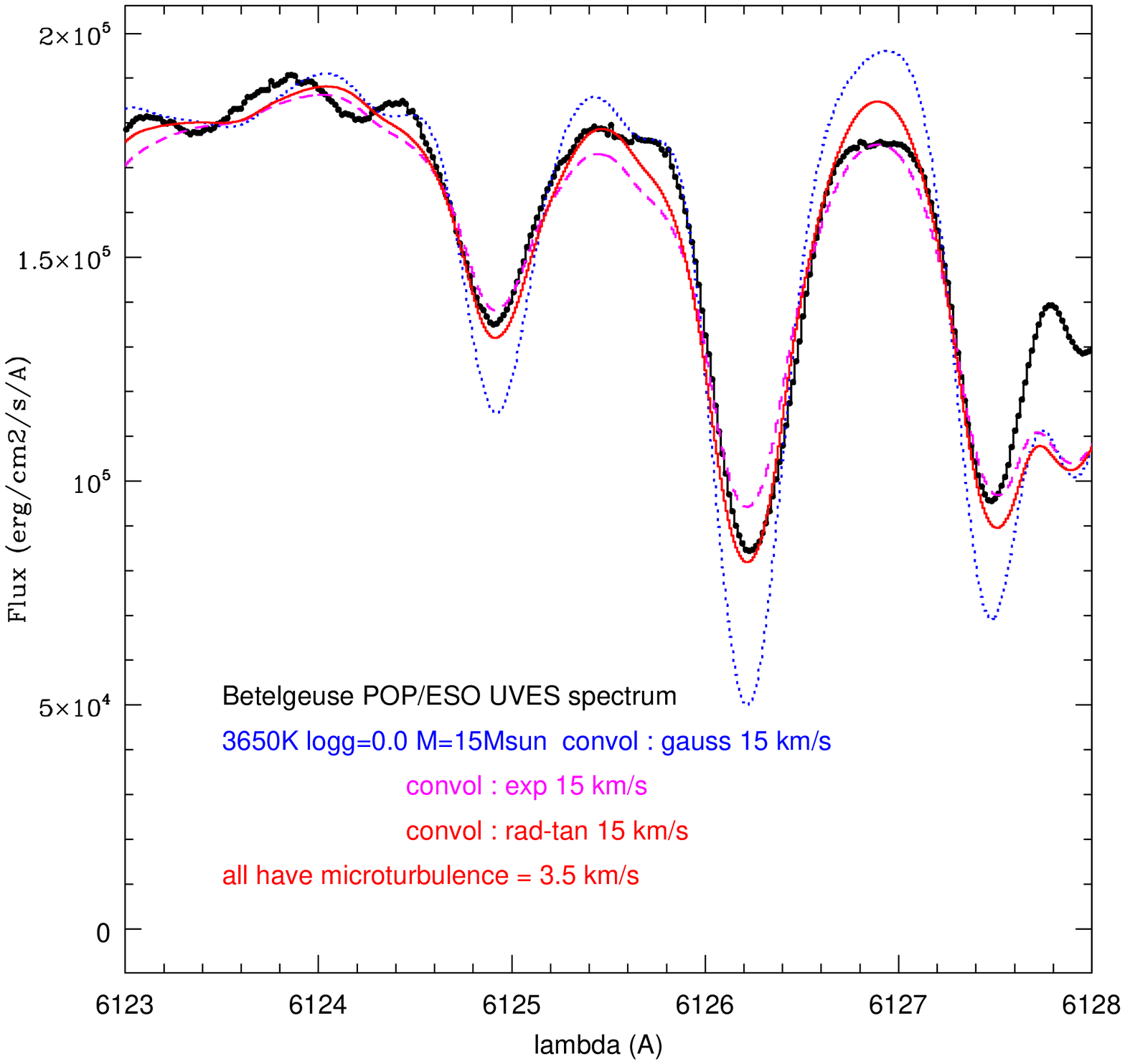}
\caption{Impact of macroturbulence on calculated spectra.
High resolution observed spectrum of Betelgeuse (black line with dots)
in the vicinity of the $gamma$' 0-0 band-head. A 
microturbulence velocity of 3.5km/s was adopted, and the spectrum is best 
matched setting a radial-tangential macroturbulence of 15km/s (thin red line).
Also shown are the effect of exponential 15km/s (magenta dashed line), and  classical Gaussian
15km/s (blue dotted line) macroturbulences.}
\end{figure}
The radial-tangential distribution best fits the data. lending support
to large granules rising in the atmosphere, with their upper layers
moving horizontally before descending vertically again between the granules.
Note that 15km/s is largely supersonic, which makes it difficult to advocate 
for such a turbulent velocity on small scales. The microturbulence 
velocity is not well constrained in RSGs but is quoted to be
of the order of 2 to 6km/s \citep[e.g.][]{2000ApJ...530..307C, 
1989ApJS...71..559L}, with the high end values being supersonic! 
Line profile studies are impeded 
by line blending in cool supergiants. The IR region is clearly a better choice
for such studies that should be intensified for a better understanding
of RSG atmospheres. 

\subsection{Non-LTE effects in the formation of molecular lines}
I give here a quick account of possible non-LTE effects in RSG
atmospheres, summarising what was exposed in \citet{2008PhST..133a4003P}.
TiO numerous electronic transition lines in the optical are a
notorious cause of heating of the outer layers of cool stars.
Surface heating or cooling only happens if the opacity is 
in absorption, and scattering has no effect. It was suggested already
by \citet{1975MNRAS.170..447H} that 
 some molecular lines do indeed form closer to scattering than pure 
absorption in the tenuous outer layers of red giants. 
Radiative rates for the optical and near IR TiO electronic transitions 
are of the order of $2.5\times10^7s^{-1}$, whereas estimates in RSG
 atmospheric conditions lead to $2\times10^3s^{-1}$ for collisions 
with electrons based on
available recipes from \citet{1962ApJ...136..906V}, and \citet{1968slf..book.....J},
and to $2\times10^6s^{-1}$ for collisions with hydrogen, based on the 
modification by 
 \citet{1993PhST...47..186L}  of the formula of \citet{1969ZPhy..228...99D}.
 Admittedly these approximations are far from justified for molecular transitions,
 and more theoretical and experimental work should be devoted to the 
determination of such 
 collisional rates [note the interesting 
 work by \citet*{Badieetal2008} on quenching rates in YO].
 Assuming radiative processes dominate over collisional ones in RSG atmospheres
 we may estimate the affect on the temperature structure. A large cooling indeed 
 occurs in surface layers if TiO transitions are supposed to occur in scattering.
 This does not lead to large changes in the TiO band strengths themselves, 
 but other lines are affected. Interestingly the cooling is what is required 
 to allow observed 12$\mu$m H$_2$O lines to be reproduced \citep{2006ApJ...637.1040R},
 but calculations with the cooled model do not show a good agreement with 
 the 12$\mu$m  spectrum, and OH lines become too strong. So, this simple way
 of modelling non-LTE effects in TiO transitions is not conclusive. 
 A full NLTE treatment of the electronic
transitions, taking into account optical depth effects, as the lines may become 
optically very thick, would be of great value. Also the coupling with 
hydrodynamics must be studied.

\section{3D Simulations of Red Supergiant Atmospheres}
I have shown above, through selected examples that classical 1D, hydrostatic,
and LTE model atmospheres, although leading to many successes \citep[see e.g.][]{2004AAS...205.1201L}, 
suffer limitations that have to be overcome. 
We know real stars are not 1D, static and in LTE!
In particular in the case of RSGs, the hydrostatic approximation must be 
abandoned, and an hydrodynamical description used instead. This is not easy nor cheap
however, as hydrodynamical equations, to be solved in 3D, must be coupled to 
the radiation field. There is an exchange of energy between the gas and the radiation
field. Huge progress has been made in the past years, and we do have a small 
number of simulations for RSGs. These are star-in-a-box calculations made with the 
Co5bold code \citep{2002AN....323..213F}. The whole star is put in the cartesian grid
of the simulation volume, with an inner central boundary condition to avoid the 
nuclear burning core. The radiation field is described using a grey opacity.
More details can be found in \citet{2008A&A...483..571F}. Current simulations have 
up to 315$^3$ points, with $T_{\rm eff}\approx$3500K, 12M$_{\odot}$, and a duration of a 
few years stellar time. Movies, and snapshots showing temperature, density, entropy,
or velocity distributions can be found at http://www.astro.uu.se/\~\ bf/.
The simulation snapshots can be used to compute detailed polychromatic
radiative transfer, which is by itself a very heavy task, and cannot 
be performed during the hydrodynamical calculations. This has been done by A. 
Chiavass during his PhD thesis \citep{andreaPHD}, available at 
http://www.graal.univ-montp2.fr/hosted/chiavassa/publi.html.
The picture that emerge is that of a granulation with size
 a little smaller than the 
stellar radius, about as predicted by \citet{1975ApJ...195..137S}, velocities 
of tens of km/s (largely supersonic), and time-scales of months to years.
The appearance of that granulation depends much on wavelength, with very great contrasts
in the optical due to TiO absorption, and the strong dependency of the Planck 
function on temperature. The granulation is much less contrasted in the IR, but 
varies in aspect between the continuum and, e.g., strong CO lines. This results
in a strong interferometric signal, both in visibility and phase, as shown by
\citet{2008arXiv0802.1403C}, that should allow a detailed characterisation 
of RSGs granulation pattern in a very near future.
Interesting results on line profiles and asymmetries are already available.
Fig. 3 shows an H$_2$O line as observed at 12.2 microns by 
\citet{2006ApJ...637.1040R} in Betelgeuse.
\begin{figure}[!ht]
\plotone{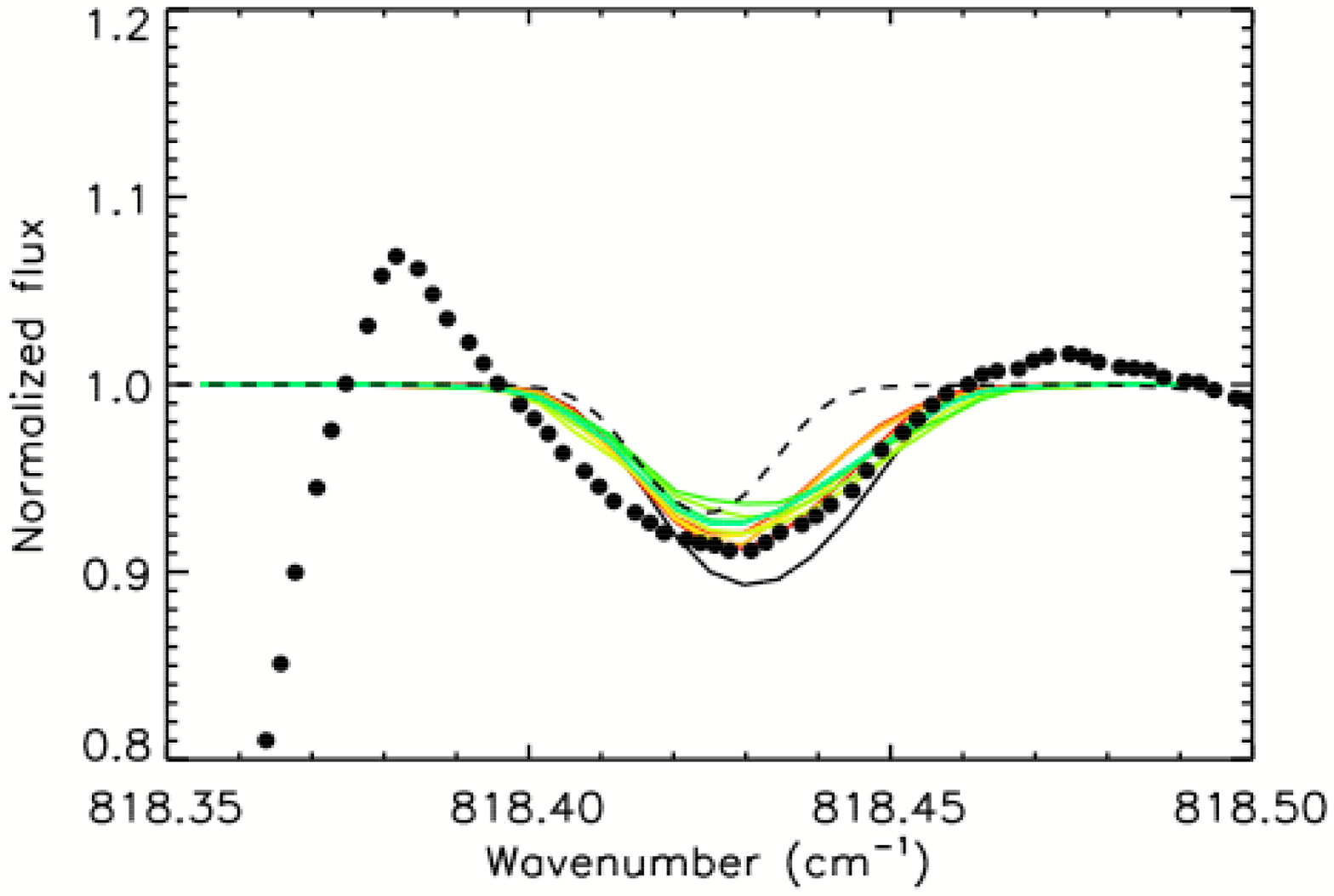}
\caption{H$_2$O line near 12$\mu$m. Dots: TEXES observation \citep{2006ApJ...637.1040R}.
Dashed line: MARCS 1D spectrum, with a microturbulence of 2km/s and an
exponential macroturbulence of 13km/s.
Thin lines: spectra for different snapshots from the same 3D simulation. The line
broadening and shifts result from the velocity field generated by the simulation. No
fudge micro- or macroturbulence  are necessary.}
\end{figure}
The 1D calculation using an hydrostatic MARCS model atmosphere, is far 
from matching the line strength or width, despite the use of
 ad'hoc micro- and macroturbulence velocities. The 3D calculations
using the velocity field provided by the simulation do show shifts, and asymmetries
that vary with time, and the line width is close to the observation, without the need
for extra macroturbulence. This gives support to the radiative-hydrodynamical
simulations. The agreement is not perfect though, as is also shown in Fig. 4 for an 
\begin{figure}[!ht]
\plotone{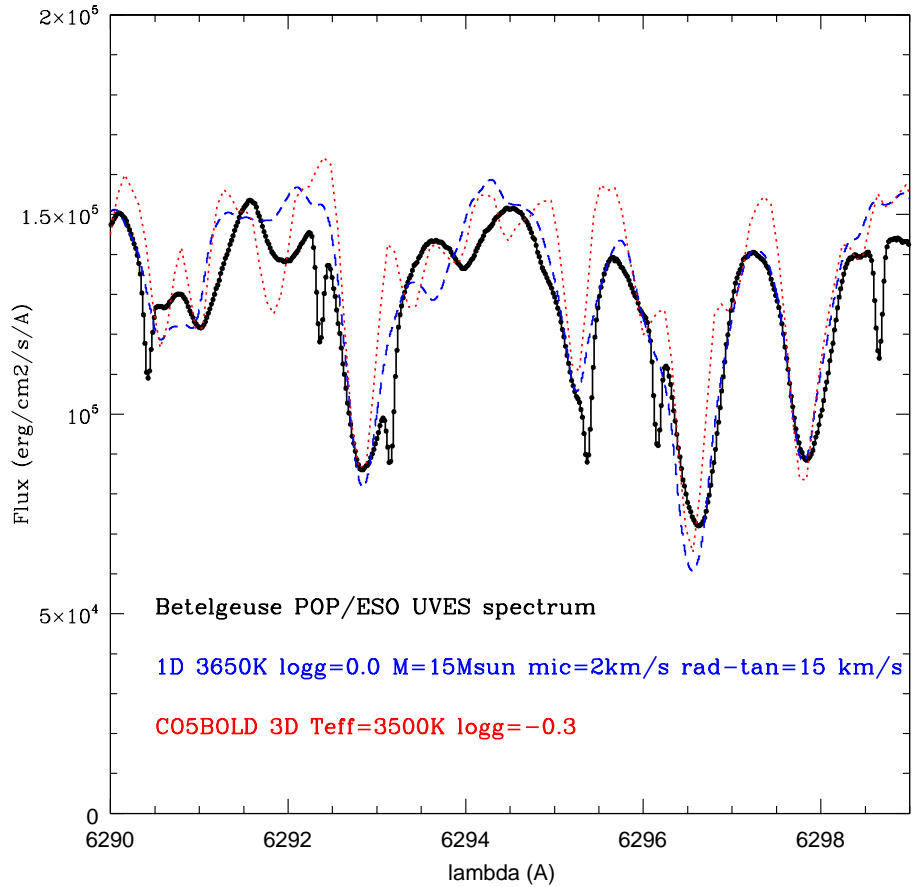}
\caption{Detail of synthetic spectra for Betelgeuse. The thick black line with dots
is the observed POP/UVES Betelgeuse spectrum. The blue dashed line is a MARCS 1D 
hydrostatic model spectrum with a microturbulence velocity of 2km/s and a 15km/s 
radial-tangential macroturbulence velocity distribution. The red dotted line is a 
Co5bold 3D hydrodynamical simulation snapshot spectrum, computed using the velocity
field of the simulation.
The adopted stellar parameters are slightly different but the main difference 
is in the velocity fields leading to line broadening : purely ad'hoc in the 1D case,
and stemming from the radiative-hydrodynamical calculation in the 3D case.}
\end{figure}
optical region of the spectrum in a TiO band. Lines appear slightly too narrow
in the 3D simulated spectrum. The 1D MARCS spectrum 
does a better job, but at the price of two fudge parameters: a microturbulence
and a macroturbulence velocity distributions (respectively gaussian and 
radial-tangential). So, the velocity dispersion of the 3D simulation seems a
little too small, but is not very far from what is observed in RSGs.
Inspection of larger chunks of spectra shows that the contrast
of TiO bands is lower in the 3D model than in the corresponding 1D hydrostatic model.
The explanation
lies probably in the fact that the temperature gradient is too shallow
in the 3D simulations. With a larger gradient in the line formation region, the contrast
between strong and weak lines, or continuum would be greater.

\section{Conclusions and Prospects}
Classical 1D, LTE, hydrostatic model atmospheres are currently easily computed
with a very detailed account of the wavelength dependence of opacity and radiation.
They also include radiation pressure computed in detail at all depths. Input 
physical data, especially line data, is now adequate for the computation of cool stars, 
including RSGs. The drawback is that convection, and turbulent pressures are
accounted through much too simple recipes. On the contrary the intricacy of 
3D geometry
with complicated velocity, temperature, and density distributions, certainly 
present in real RSGs, can only be described in 3D radiative-hydrodynamical simulations.
This is of course possible only at the expense of simplifications 
in the treatment of the radiation field and related quantities, that 
 are computed in the grey approximation (using Rosseland 
or Planck mean opacities). The 3D models do then predict velocity fields and temperature 
inhomogeneities, and detailed non-grey radiative transfer can be calculated 
a posteriori to produce images, and spectra. However, the velocity dispersion
and the temperature gradient seem too shallow, when compared to observations.
The models must be further developed, in particular with a non-grey radiative
transfer, based on a small number of opacity bins, the only tractable solution for now.
This should lead to greater temperature gradients. The inclusion of 
radiative pressure could help increase velocity fields to the observed levels.
Once these 3D models are in a satisfactory state, validated by observations, 
we should devise recipes that can be incorporated in easier to compute, and 
manipulate, 1D (or 2D) models. 

\acknowledgements 
This reflects the work of a large number of people, esp. for the MARCS model atmospheres: Bengt Gustafsson, Bengt Edvardsson, Kjell Eriksson, and others; 
for the 3D models: Andrea Chiavassa, Bernd Freytag, Hans-G\"unter Ludwig, Matthias Steffen; and for the observations and analyses, Phil Massey, Emily Levesque, Eric Josselin, Knut Olsen, David Silva, and Geoff Clayton.
This work has benefited of discussions with Eric Josselin who also contributed with some computations. Andrea Chiavassa kindly provided some illustrations.
I wish to thank the organisers for inviting me, and for financial support.
This research was supported in part by
the French National Research Agency (ANR) through program number
ANR-06-BLAN-0105.


\end{document}